\documentclass[prl,showpacs,twocolumn,amsfonts]{revtex4}

\usepackage{graphicx}% Include figure files
\usepackage{bm}% bold math
\usepackage{chemarrow}
\usepackage{stackrel}

\begin{document}

\title{A constrained random-force model for weakly bending semiflexible polymers}
\author{Panayotis Benetatos and Eugene M. Terentjev}
\affiliation{Cavendish Laboratory, University of Cambridge, J. J.
  Thomson Avenue, Cambridge, CB3 0HE, United Kingdom }

\date{\today}

\begin{abstract}
The random-force (Larkin) model of a directed elastic string subject to
quenched random forces in the transverse directions has been a
paradigm in the statistical physics of disordered systems.  In this
brief note, we investigate a modified version of the above model where the total
transverse force along the polymer contour and the related total torque, in each realization of
disorder, vanish. We discuss the merits of adding these constraints
and show that they leave the qualitative
behavior in the strong stretching regime unchanged, but they reduce the
effects of the random force by significant numerical
prefactors. We also show that a transverse random force
  effectively makes the filament softer to compression by inducing
  undulations. We calculate the related linear compression coefficient
  in both the usual and the constrained random force model.
\end{abstract}
\maketitle

The random-force model was introduced by Larkin
as a short-distance approximation to study the effect of a quenched random
potential 
on the Abrikosov lattice \cite{Larkin,Blatter_review}. Because of its simplicity (the
relevant functional integrals are Gaussian), it admits exact solutions
and has become a paradigm in the physics of disordered elastic
manifolds \cite{Blatter_new}. In a recent publication \cite{pap1}, we
used it to analytically investigate the effect of a quenched disordered
environment on a strongly stretched wormlike chain. Of course, the
randomness that biopolymers are subject to in the cellular environment
is much more complicated \cite{Kumar1,Kumar2}, but this model yields
some very simple analytical results for the weakly bending case.

\begin{figure}
\includegraphics[angle=0,width=0.4\textwidth]{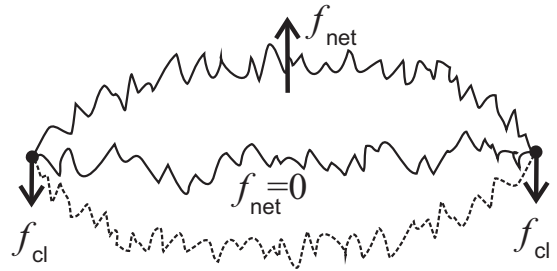}
\caption{An unconstrained random linear force density includes realizations with a
  finite net transverse force balanced by the clamps. Such a net force
  yields an overall bend in addition to the microscopic undulations.\label{clamps}}
\end{figure}

The random-force model is mathematically well-defined and its analysis
is valid. However, its relevance to experimental measurements is
obscured by the fact that the distribution of quenched disorder
includes realizations of the random force which have non-zero total
force along the polymer contour. The total transverse force in any
given 
experiment is either zero or non-zero and, irrespective of the size of the
polymer, these two distinct possibilities persist and do not
average. We can show, using the central limit theorem, that the
variance of the total transverse force, in the limit of long contour
length $L$, scales as $\sim L$. This is a manifestation of 
lack of self-averaging, and the disorder-averaged
value of an 
observable  is not expected to coincide with the outcome of a single
measurement in a long filament. Moreover, we should consider two
  distinct scenarios in a stretching experiment. In the first, the end points of the polymer are
  fixed in space and absorb (balance) both the total force and torque that
  the polymer feels. A net transverse random force will have a
  macroscopically manifest effect on the conformation of the
  polymer, as shown in Fig. \ref{clamps}. It will also be experienced
  by the end-point clamps. The same is true
  for a net torque. In the second scenario, the polymer ends are free
  to equilibrate in the random potential and the net force and torque
  identically vanish. Averaging over all possible random force
  realizations could be misleading in the case of vanishing net force or
  torque. These observations motivate us to study a modified
version of the random-force model which includes the constraint of
vanishing total force along the polymer contour in every single realization of the
quenched disorder \cite{J-P}. We also consider the additional constraint
prescribed by the condition of mechanical equilibrium that the total
torque on the polymer in every single realization of the
quenched disorder vanishes. The only effect of such a random force is
the deformation of the polymer through random undulations. A quenched
transverse random force could also result from the interaction of a
stretched semiflexible polyampholyte with a transverse electric field
(assuming screened intrapolymer interactions). It is known that the
constraint of global neutrality may affect the behavior of a
polyampholyte \cite{Yamakov}. This provides further motivation for
studying the constrained random force model.

%%%%%%%%%%%%%%%%%%%%%%%%%%%%%%%%%%%%%%%%%%%%%%%%%%%%%%%
%%%%%%%%%%%%%%%%%%%%%%%%%%%%%%%%%%%%%%%%%%%%%%%%%%%%%%%%
%%%%%%%%%%%%%%%%%%%%%%%%%%%%%%%%%%%%%%%%%%%%%%%%%%%%%%%%%%%%%%%%%%%%

We condider a weakly bending wormlike chain of total contour length $L$ in $1+1$
dimensions parametrized by the displacement $y(s)$ perpendicular to
the aligning direction ($x$), where $s$ is the arc-length along the polymer
contour. An axial force pair $f$ is applied at its end-points in the
$x$ direction and a spatially varying
random linear force density $g(s)$ is applied along the polymer
contour in the transverse direction. The elastic energy functional is
given by 
\begin{eqnarray}
\label{H_g}
{\cal H}_g[y(s)]&= &\frac{\kappa}{2}\int_0^L ds\: \Big(\frac{\partial^2
  y(s)}{\partial s^2}\Big)^2 -  \int_0^L ds\: g(s)y(s)\nonumber\\
& &+\frac{1}{2}f\int_0^L
  ds\: \Big(\frac{\partial
  y(s)}{\partial s}\Big)^2 -fL\;,
\end{eqnarray}
where $\kappa$ is the bending rigidity related to the persistence
length $L_p$
via $\kappa=\frac{1}{2}L_p k_B T$. $f>0$ corresponds to stretching
and $f<0$ corresponds to compression along the filament axis. The first term in the r.h.s. of
Eq. (\ref{H_g}) penalizes bending, the second term expresses the
interaction with the transverse random force, and the remaining two
express the interaction with the stretching force ($-f[x(L)-x(0)]$). $g(s)$ acts as quenched
disorder. Its probability distribution is given by 
\begin{eqnarray}
\label{Pofc}
{\cal P}[g(s)] &\sim&
\int_{-\infty}^{+\infty}d{\lambda}\:\exp\Big[i\lambda\int_0^Lds\:
g(s)\Big]\nonumber\\
& &\times \int_{-\infty}^{+\infty}d{\mu}\:\exp\Big[i\mu\int_0^Lds\:
s\: g(s)\Big]\nonumber\\
& &\times \exp\Big[       -\int_0^L ds\:
\frac{\big(g(s)\big)^2 }{2{\Delta}_g} \Big]\;.
\end{eqnarray}
This is Gaussian at any arc-length position $s$ with variance ${\Delta}_g$,
modified by the constraint of zero total force which is expressed by the integral over
$\lambda$ using the
Fourier representation of the Dirac $\delta$-function, and  the
constraint of zero total torque which is expressed by the integral over
$\mu$. For the sake of simplicity, we consider hinged-hinged boundary
conditions:
\begin{eqnarray}
\label{bc}
y(0)&=&y(L)=0\nonumber\\
{\partial_s^2 y(s)}|_{s=0}&=&{\partial_s^2 y(s)}|_{s=L}=0\;.
\end{eqnarray}

%%%%%%%%%%%%%%%%%%%%%%%%%%%%%%%%%%%%%%%%%%%%%%%%%%%%%%%%%%%%%%%%%%%%%

The force-extension relationship is obtained from
\begin{eqnarray}
\label{ave(x)}
\overline{\langle x(L) \rangle} &=& L-\frac{1}{2}\int_0^L ds\:
\overline{\Big \langle \Big(\frac{\partial y(s)}{\partial s}\Big)^2
  \Big \rangle}\:,
\end{eqnarray}
where $\langle...\rangle$ and $\overline{(...)}$ denote
thermal and disorder averages, respectively.  The
boundary conditions of Eq. (\ref{bc}) allow us to to expand $y(s)$ in
a series of sines:
\begin{eqnarray}
y(s)=\sum_{l=1}^{\infty} A^{(l)}\sin\Big(\frac{l\pi}{L}s\Big)\;.
\end{eqnarray}
We calculate disorder-averaged correlators using the replica method as in Ref. \cite{pap1}:
\begin{eqnarray}
\overline{\langle A^{(l)}A^{(m)}\rangle}={\displaystyle
  \lim_{n\rightarrow 0}}\frac{k_BT}{n}\sum_{a=1}^n(C^{-1})^{(lm)}_{aa}\;,
\end{eqnarray}
where
\begin{eqnarray}
\label{C}
& &C_{ab}^{(lm)}={{\chi}_l}^{-1}\delta_{ab}\delta_{lm}-\frac{L{\Delta}_g}{2k_B
  T}\Big(\delta_{lm}-\frac{8}{\pi^2}\frac{1}{lm}\nonumber\\
& &-\frac{8}{\pi^2}\frac{(-1)^{l+m}-(-1)^l+2(-1)^m}{lm}\Big){\bf
1}_{ab}\,,
\end{eqnarray}
\begin{eqnarray}
\chi_l=\frac{2}{L\Big(\kappa\big({l\pi}/{L}\big)^4+f\big({l\pi}/{L}\big)^2\Big)}\;,
\end{eqnarray}
${\bf 1}_{ab}$ being an $n\times n$ matrix with all of its elements
equal to $1$, and 
\begin{eqnarray}
\label{Cmatrix}
& &(C^{-1})_{ab}^{(lm)}\;\stackbin[{\scriptscriptstyle n \rightarrow 0}]{}{\longrightarrow}\;\delta_{lm}\Big(\chi_l\delta_{ab}+\chi_l^2\frac{L
  \Delta_g}{2 k_BT}{\bf 1}_{ab}\Big)-\frac{4L \Delta_g}{\pi^2
  k_BT}\Big(\frac{\chi_l \chi_m}{lm}\nonumber\\
& &+\chi_l\chi_m\frac{(-1)^{l+m}-(-1)^l+2(-1)^m}{lm}\Big){\bf 1}_{ab}\;.
\end{eqnarray}
The first term in the r.h.s. of Eq. (\ref{Cmatrix}) is related to the
usual random-force model whereas the remaining terms express the
contribution of the constraints.

%%%%%%%%%%%%%%%%%%%%%%%%%%%%%%%%%%%%%%%%%%%%%%%%%%%%%%%%%%%%%%%%%%%%%
In the absence of the zero-torque constraint (that is, without
  the second factor in the rhs of 
Eq. (\ref{Pofc})), Eqs. (\ref{C}) and (\ref{Cmatrix}) become 
\begin{eqnarray}
\label{C1}
& &C_{ab}^{(lm)}={{\chi}_l}^{-1}\delta_{ab}\delta_{lm}-\frac{L{\Delta}_g}{2k_B
  T}\Big(\delta_{lm}\nonumber\\
& &-\frac{2}{\pi^2}\frac{(-1+(-1)^l)(-1+(-1)^m)}{lm}\Big){\bf
1}_{ab}
\end{eqnarray}
and 
\begin{eqnarray}
\label{Cmatrix1}
& &(C^{-1})_{ab}^{(lm)}\;\stackbin[{\scriptscriptstyle n \rightarrow 0}]{}{\longrightarrow}\;\delta_{lm}\Big(\chi_l\delta_{ab}+\chi_l^2\frac{L
  \Delta_g}{2 k_BT}{\bf 1}_{ab}\Big)\nonumber\\
& &-\frac{L{\Delta}_g}{\pi^2 k_B T}\chi_l\chi_m\frac{(-1+(-1)^l)(-1+(-1)^m)}{lm}{\bf 1}_{ab}\;,
\end{eqnarray}
respectively.

%%%%%%%%%%%%%%%%%%%%%%%%%%%%%%%%%%%%%%%%%%%%%%%%%%%%%%%%%%%%%%%%%%%%%5

In the strong stretching regime which is defined by $f\gg \kappa/
L_p^2$ for filaments with $L \gg L_P$ or $f\gg \kappa /L^2$ for
filaments with $L_p \gg L$, we obtain:
\begin{eqnarray}
\label{fext}
& &\overline{\langle x(L)
  \rangle}=L-\frac{k_BT L}{4 \sqrt{\kappa
    f}}-\frac{1}{30}\frac{\Delta_g L^2}{f^2} \;.
\end{eqnarray}
The last term in the r.h.s. of the above equation is the decrease in
the end-to-end distance due to the undulations caused by the
transverse random force. In the usual (unconstrained) random-force
model, the numerical prefactor of that term is $1/12$ \cite{footnote}.  The
zero-total-force constraint merely reduces this prefactor  to $1/24$, and adding the
zero-total-torque constraint further reduces it to $1/30$.

We also investigate the effect of the constraints on the pulling-force
response of the width of the transverse undulations (mean square
transverse displacement at the mid-point) of the stretched
wormlike chain. In the strong stretching regime, we obtain:
\begin{eqnarray}
\label{transv}
& &\overline{\Big\langle \Big(y(s=\frac{L}{2})\Big)^2
  \Big\rangle}=\frac{1}{4}\frac{L k_B T}{f}+\frac{1}{96}\frac{\Delta_g L^3}{f^{2}}\:.
\end{eqnarray}
In a similar fashion as with the force-extension response, the
zero-total-force constraint, reduces the effect of the
random force, which is expressed by the last term, by a factor of
$1/2$ (it changes the numerical prefactor from $1/48$ to $1/96$). The zero-total-torque constraint
does not affect the width of transverse fluctuations. We point out, however, that these
simple results (Eqs. (\ref{fext}) and (\ref{transv})) only hold in the
strong stretching regime. For arbitrary stretching force $f$, the
difference between the ususal and the constrained random force model
is more complicated and depends on the system size ($L$) as can be seen from Eq. (\ref{Cmatrix}).

%%%%%%%%%%%%%%%%%%%%%%%%%%%%%%%%%%%%%%%%%%%%%%%%%%%%%%%%%%%%%%%%%%%%%%%

A wormlike chain at zero temperature does not yield to a small axial
compressional force below the critical buckling value
\cite{Euler,Landau}. At any finite temperature, however, thermally
induced undulations smooth out the buckling and give rise to a {\em linear}
response for small compressional forces \cite{KroyFrey}. A quenched disordered
environment induces similar undulations and therefore a similar
contribution to the linear response coefficient. We consider the constrained
random-force as a simple model for the quenched disordered environment. Assuming $f<0$
(compression) and $L_p\gg L$, we obtain the average projected length of the filament in the
direction of the compressing force to leading (linear) order in $|f|$:
\begin{eqnarray}
\label{compr}
& &\overline{\langle x(L)
  \rangle}=L-\frac{1}{12}\frac{k_BTL}{\kappa}-\frac{1}{180}\frac{k_BTL^4}{\kappa^2}|f|\nonumber\\
& &-\frac{31}{302400}\frac{\Delta_g L^6}{\kappa^2}-\frac{67}{3326400} \frac{\Delta_g L^8}{\kappa^3}|f|\:.
\end{eqnarray}
The first line Eq. (\ref{compr}) expresses the contribution of the
thermally induced undulations whereas the second reflects the effect
of the
transverse random force. Both contributions modify the classical Euler
behavior of an elastic rod before buckling. 
This second line becomes 
\begin{eqnarray*}
-\frac{1}{1890}\frac{\Delta_g L^6}{\kappa^2}-\frac{1}{9450}
\frac{\Delta_g L^8}{\kappa^3}|f| 
\end{eqnarray*}
 in the usual
(unconstrained) random force model and 
\begin{eqnarray*}
-\frac{13}{120960}\frac{\Delta_g L^6}{\kappa^2}-\frac{37}{1814400}
\frac{\Delta_g L^8}{\kappa^3}|f|,
\end{eqnarray*}
if we add the
zero-total-force constraint. In contrast to the non-linear stretching response (Eq. (\ref{fext})), where the
contribution of the random-force undulations decreases relative to
that of their thermal counterpart as the pulling force increases, in
the linear compressional response the the two contributions enter on equal
footing. We point out that the thermal and the quenched-disorder
contributions differ qualitatively
as far as their dependence on the filament parameters (bending
stiffness $\kappa$, and total contour length $L$) is concerned. This
difference could in principle be useful to probe the existence of
non-thermal lateral random forces on compressed microtubules in the
cellular environment. For $f=0$ and $T=0$, a filament of fixed bending stiffness tends to
crumple under the load of
the random forces as its size $L$ increases. This is typical of the
Larkin model and it remains unaffected by the zero-total-force
and zero-total-torque constraints. Of course, within the context of our weakly-bending
approximation we can only identify a tendency towards crumpling and
not an actual transition.

%%%%%%%%%%%%%%%%%%%%%%%%%%%%%%%%%%%%%%%%%%%%%%%%%%%%%%%%%%%%%%%%%%%%%%

In this brief note, we have investigated a modification of the
random-force model which excludes realizations of disorder with a
non-vanishing total force and a non-vanishing total torque. These are 
non-self-averaging contributions. We have calculated the force-extension
response of a wormlike chain and the width of its transverse
undulations in the strong stretching regime 
using this modified model. We have also calculated the linear
compressional response using both the usual and the constrained random
force model. We have shown that the
constraints leave the calculated behavior qualitatively unchanged but
they reduce the effect of the random forces by significant numerical
prefactors. The dependence of the disorder-induced contribution on
the size of the system ($L$), which grows in the thermodynamic
limit, suggests that the random force model, even
in its modified version, is inherently non-self-averaging. It would be
interesting to calculate how the constraints discussed here would
modify the free energy distribution functions analyzed in Ref. \cite{Blatter_new}. Although in this work we
deal with a weakly bending wormlike chain, our analysis of the vanishing
total force and torque constraints also holds in other applications of the random-force model
(e.g., flux lines in type II superconductors or liquid crystals in
random media \cite{Pelcovits}).

This work was based on discussions held at the KITPC (Chinese Academy of Sciences Grant
No. KJCX2.YW.W10); it was also supported by EPSRC via the University of Cambridge TCM Programme Grant.


\begin{thebibliography}{}
\bibitem{footnote}
A discrepancy by a factor of 2 with the corresponding term in Ref. \cite{pap1} is due
to a typo in Ref. \cite{pap1}.
\end{thebibliography}
\end{document}